\documentclass[a4paper,fleqn,usenatbib]{mnras}

\usepackage{epsfig}
\usepackage[a4paper]{geometry}
\usepackage{amsmath}
\usepackage[T1]{fontenc}
\usepackage{ae,aecompl}
\usepackage{graphicx}	
\usepackage{amssymb}	
\usepackage{color}
\usepackage[utf8]{inputenc}
\usepackage{ragged2e}
\newcommand{\Msun}{$\mathrm{M}\odot$}

 \usepackage{tabularx}
\usepackage{footnote}
\usepackage{subcaption}
\usepackage{afterpage}
\usepackage{verbatim}
\newcommand{\psrchive}{\texttt{psrchive}}

\newcommand{\tempo}{\texttt{TEMPO}}
\newcommand{\tempotwo}{\texttt{TEMPO2}}
\newcommand{\nemodel}{\texttt{NE2001}}
\title[Two MSPs from HTRU-North]
{The discovery of two mildly-recycled binary pulsars in the Northern High Time Resolution Universe pulsar survey}

\author[Berezina et al.]
{M.~Berezina$^{1}$\thanks{E-mail: mberezina@mpifr-bonn.mpg.de},
 D.~J.~Champion$^{1}$, P.~C.~C.~Freire$^{1}$, T.~M.~Tauris$^{1,2}$,  M.~Kramer$^{1,3}$,
 \newauthor
 A.~G.~Lyne$^{3}$, B.~W.~Stappers$^{3}$, L.~Guillemot$^{1,4,5}$, I.~Cognard$^{4,5}$, E.~D.~Barr$^{1}$,
\newauthor
  R.~P.~Eatough$^{1}$, R. Karuppusamy$^{1}$, L.~G.~Spitler$^{1}$, G.~Desvignes$^{1}$
\\
$^{1}$Max-Planck-Institut für Radioastronomie, Auf dem Hügel 69, D-53121 Bonn, Germany\\
$^{2}$Argelander-Institut f\"ur Astronomie, Universit\"at Bonn, Auf dem H\"ugel 71, 53121 Bonn, Germany\\
$^{3}$Jodrell Bank Centre for Astrophysics, School of Physics and Astronomy, The University of
    Manchester, Manchester M13 9PL, UK\\    
$^{4}$Laboratoire de Physique et Chimie de l'Environnement et de l'Espace, LPC2E, CNRS-Universit\'{e} d'Orl\'{e}ans,
F-45071 Orl\'{e}ans, France\\
$^{5}$Station de Radioastronomie de Nan\c{c}ay, Observatoire de Paris, CNRS/INSU, F-18330 Nan\c{c}ay, France\\
}

\date{Accepted XXX. Received YYY; in original form ZZZ}

\pubyear{2017}

\begin{document}
\label{firstpage}
\pagerange{\pageref{firstpage}--\pageref{lastpage}}
\maketitle

\begin{abstract}

We report the discovery and the results of follow-up timing observations of PSR J2045+3633 and PSR J2053+4650, two binary pulsars found in the Northern
High Time Resolution Universe pulsar survey being carried out with the Effelsberg radio telescope.
Having spin periods of 31.7\,ms and 12.6\,ms respectively, and both with massive white dwarf companions, $M_{c}\, > \, 0.8\, M_{\odot}$,
the pulsars can be classified as mildly recycled. PSR J2045+3633 is remarkable due to its orbital period (32.3 days) and eccentricity $e\, = \, 0.01721244(5)$ which is 
among the largest ever measured for this class. After almost two years of timing the large eccentricity has allowed  the measurement of 
 the rate of advance of  periastron  at the 5-$\sigma$ level, 0.0010(2)$^\circ~\rm yr^{-1}$. 
Combining this with a detection of the orthometric amplitude of the Shapiro delay, we obtained the following constraints 
on the component masses (within general relativity): $M_{p}\, = \, 1.33^{+0.30}_{-0.28}\, M_{\odot}$, and $M_{c}\, = \, 0.94^{+0.14}_{-0.13}\, M_{\odot}$. 
PSR J2053+4650 has a 2.45-day circular orbit inclined to the plane of the sky at an angle $i\, = \, 85.0^{+0.8}_{-0.9}\,{\rm deg}$. In this nearly edge-on case the masses can be obtained from  the
Shapiro delay alone. Our timing observations resulted in a significant detection of this effect giving: $M_{p}\, = \, 1.40^{+0.21}_{-0.18}\, M_{\odot}$, and $M_{c}\, = \, 0.86^{+0.07}_{-0.06}\, M_{\odot}$.

\end{abstract}




\section{Introduction}

The discovery rate of binary pulsars has been rapidly increasing over the last decade
and the population currently includes $\sim\!250$ systems \citep{mhth05}.
There is a large diversity among the nature of the companion stars and the characteristics of
both the pulsars and their orbits \citep[e.g.][]{tau11}.
The companion stars detected so far are either non-degenerate main-sequence stars, semi-degenerate (and hydrogen rich) dwarfs, 
helium white dwarfs (He~WDs), carbon-oxygen (CO) or oxygen-neon-magnesium (ONeMg) WDs, or neutron stars (NSs).
Some pulsars are found with planets \citep[e.g. PSR~B1257+12,][]{wf92}, and some are members of a triple 
system \citep[e.g. PSR~J0337+1715,][]{rsa+14}.

The vast majority of the observed binary radio pulsars have been recycled via accretion of mass and angular momentum
from a companion star \citep{bv91,tv06}, often leading to formation of a millisecond pulsar (MSP). 
Accreting pulsars are observable in  X-rays \citep{bcc+97} as  low- (LMXB), intermediate- (IMXB) or high-mass X-ray binaries (HMXB), depending on the
mass of the donor star. The initial mass of the donor star has an impact on  the duration and stability of the mass-transfer phase and, hence, the efficiency
of the recycling process. This, in turn, determines the main properties of the recycled pulsar, in particular its spin period and spin-down rate, as well as the final orbital configuration.
This can clearly be seen when examining these properties as a function of the compact companion type in the known MSP population \citep{tlk12}. (In the following, we discard binary pulsars observed in dense environments
like globular clusters since these binaries are possibly formed via
exchange encounters \citep[e.g.][]{vf14} and therefore have a different formation history
compared to binary pulsars in the Galactic disk.)
For example, MSPs with low-mass He~WD companions are often fully recycled (with spin periods, $P<10\;{\rm ms}$, and
period derivatives, $\dot{P}\la 10^{-20}$), whereas MSPs with massive CO/ONeMg~WDs are often
only mildly recycled with $10 < P < 100\;{\rm ms}$ and $10^{-20} < \dot{P} < 10^{-18}$. 
The most massive donor stars are found in HMXBs.  If such systems remain bound after the second supernova (SN) explosion, they produce double NS systems.
 In wide-orbit HMXBs, the effective mass-transfer phase is so short
 that the first born NS only becomes a marginally recycled pulsar \citep{tlp15}, in some cases with a spin period 
exceeding 100~ms. A prime example of such a system is the double NS system PSR~J1930$-$1852 \citep{srm+15} which hosts a mildly recycled pulsar with $P = 185\;{\rm ms}$ and has an orbital period of 45 days.

Pulsars with a compact star companion represent the end-point of binary stellar evolution. Thus, we can use the observed
characteristics of these systems as fossil records to learn about stellar evolution and binary interactions in
their progenitor systems \citep{ltk+14}.
MSPs with He~WD companions, and which have been recycled via stable Roche-lobe overflow (RLO) in LMXBs,
possess a unique correlation between the mass of the WD and the orbital period \citep[e.g.][]{ts99,2014Istrate}.
Furthermore, for these systems the orbital eccentricities are also correlated with the orbital period \citep{phi92,pk94}, such that binaries with short orbital periods are more circular in general.
For double NS systems, it is even possible to put constraints on the properties of the second SN explosion \citep{wkk00,fsk+13,fsk+14,tlp15}.

Recycled pulsars with CO/ONeMg~WD companions (also known as intermediate-mass binary pulsars, IMBPs) were first recognized as a separate class by \citet{cam96}.
Initially, it was thought that all such binaries form via common-envelope (CE) evolution. This idea was based
on the formation scenario for PSR~J2145$-$0750 \citep{bhl+94} which was put forward by \citet{vdh94}.
However, it was later demonstrated \citep{tvs00} that such intermediate-mass binary pulsar systems
can also be formed without the need for CE evolution, in cases when the observed orbital period is larger than 3~days.
For these wider systems, the formation process was most likely dynamically stable RLO in an IMXB.

Although some consensus in our understanding of binary pulsars is emerging, it is important to keep
finding new systems that will challenge current ideas and bring forward new lines of research.
The discovery of PSR~J1614$-$2230 \citep{dpr+10}, which is the first example of a fully recycled MSP with a CO~WD companion
-- and the first NS with a precisely measured mass close to $2.0\;M_{\odot}$ -- came somewhat as a surprise.
 Subsequently, detailed modelling by \citet{lrp+11} and \citet{tlk11}, suggested that this system is the first
known example of an IMXB system which produced a radio pulsar evolving via Case~A RLO (i.e.\ mass transfer
initiated while the donor star is still burning hydrogen in its core). Hence, a third possibility 
for producing a recycled pulsar with a CO/ONeMg~WD is now accepted.
 
The number of pulsar binaries with CO/ONeMg~WDs now exceeds 30 systems. Their orbital eccentricities can vary
from the order of $10^{-6}$ to $10^{-2}$. This large spread in eccentricity suggests different formation paths. 
With only a small number  of such systems known with precisely measured masses of the stellar components 
(apart from: PSR~J1614$-$2230 \citep{rsa+14}, PSR~J1802$-$2124 \citep{fsk+10} and PSR~J0621+1002 \citep{sna+02}),
it is difficult to speculate more precisely about their previous evolution.
The discovery of new NSs with massive WD companions, which allow for mass determinations, would help shed new light on 
the formation process. Furthermore, precise knowledge of the present pulsar mass -- in combination with other 
system parameters available from pulsar timing and theoretically computed binary models -- will make it possible
to estimate the amount of accreted material (and hence to infer the birth mass of the NS), determine which binary interactions were at work,
and, consequently, try to deduce the formation history of the system as a whole. 
As an additional bonus, precise mass measurements can also serve for constraining the equation-of-state of cold dense matter
within NSs \citep{2016ARA&A..54..401O}.

Here, we present the discovery and follow-up timing of two new mildly recycled MSPs which appear to be promising for the 
aforementioned purposes. They were discovered in June~2014 in the Northern High Time Resolution Universe 
survey \citep{bck+13}. The structure of the paper is as follows: In Section~2, we introduce the survey and the
discoveries; in Section 3, we describe our timing campaigns; in Section~4, we present the results of timing and polarisation analysis, discuss the details
of the performed mass measurements and speculate about the eccentricity-binary period relation for IMBPs, 
and, finally, in Section~5, we summarize our work.

\section{MSP discoveries in the HTRU-North Survey}

\subsection{HTRU-North}
The Northern High Time Resolution Universe survey for pulsars and fast transients is being conducted with  the 100-m Effelsberg radio telescope in Germany.  
It makes use of the 21-cm seven-pixel  multibeam receiver and the  polyphase filterbank backend providing a time resolution of  54 $\mu$s 
and a 300-MHz passband centered around 1.36 GHz and split into 512 channels.

Being a counterpart of the Southern survey  \citep[HTRU-South;][]{kjv+10}, it follows the same  observing convention with the sky split into 
three regions (the only difference is the integration time):  low Galactic latitudes - $|b|<3.5^{\circ}$ 
where each sky pointing is observed for 25 minutes, medium Galactic latitudes - $|b|<15^{\circ}$ with the integration time of 3 minutes
and high Galactic latitudes - $|b|>15^{\circ}$ with 1.5 minute integrations. Pointings located in and near the Galactic plane are expected 
to harbour the highest yield of  interesting PSR-WD or DNS systems  \citep[or even exotic systems with black hole companions; e.g.][]{Belczynski2002}
that can be used for such scientific purposes as studying stellar evolution, testing general relativity,
constraining the equation-of-state of supra-nuclear matter \citep[e.g.][]{lrr-2008-8}.  For this reason, we are currently concentrating  
on the mid-latitude region with 3-minute integration times  as a shallow survey of the  plane  can speed up the discovery and further studies of the brightest pulsars.

The northern part of the plane  (as well as the whole northern sky) visible from Effelsberg has not yet been investigated in the L-Band,
except for some regions of overlap with other 21-cm surveys like PALFA \citep[e.g.][]{cfl+06,laz13},
whose declination range is limited leaving large areas of the Galaxy uncovered, or with the southern surveys
(HTRU-South, PMPS - see, for example, \citep{lfl+06}). A  high observing frequency reduces the impact of interstellar medium effects (such as dispersion and scattering)
 allowing the Galactic plane to be probed more deeply for
 distant objects. This depth of scanning distinguishes the HTRU-North from and, at the same time,
  makes it complementary to the low-frequency Northern-sky surveys such as the 350-MHz GBNCC  \citep{slr+14}, GBT drift-scan  
 \citep{karako2015} or the 140-MHz LOFAR pulsar surveys \citep{clhs+14} and the all-sky Arecibo 327 MHz drift survey \citep{2013ApJ...775...51D} 
which use the steep spectral index observed in most pulsars to detect  nearby weak sources.
The full description of the survey and previous discoveries can be found in \citet{bck+13, 2017MNRAS.465.1711B}.

\begin{figure*}
\begin{subfigure}{.5\textwidth}
 \centering
  \includegraphics[height = 7.1 cm,  angle= 0, scale = 1.07]{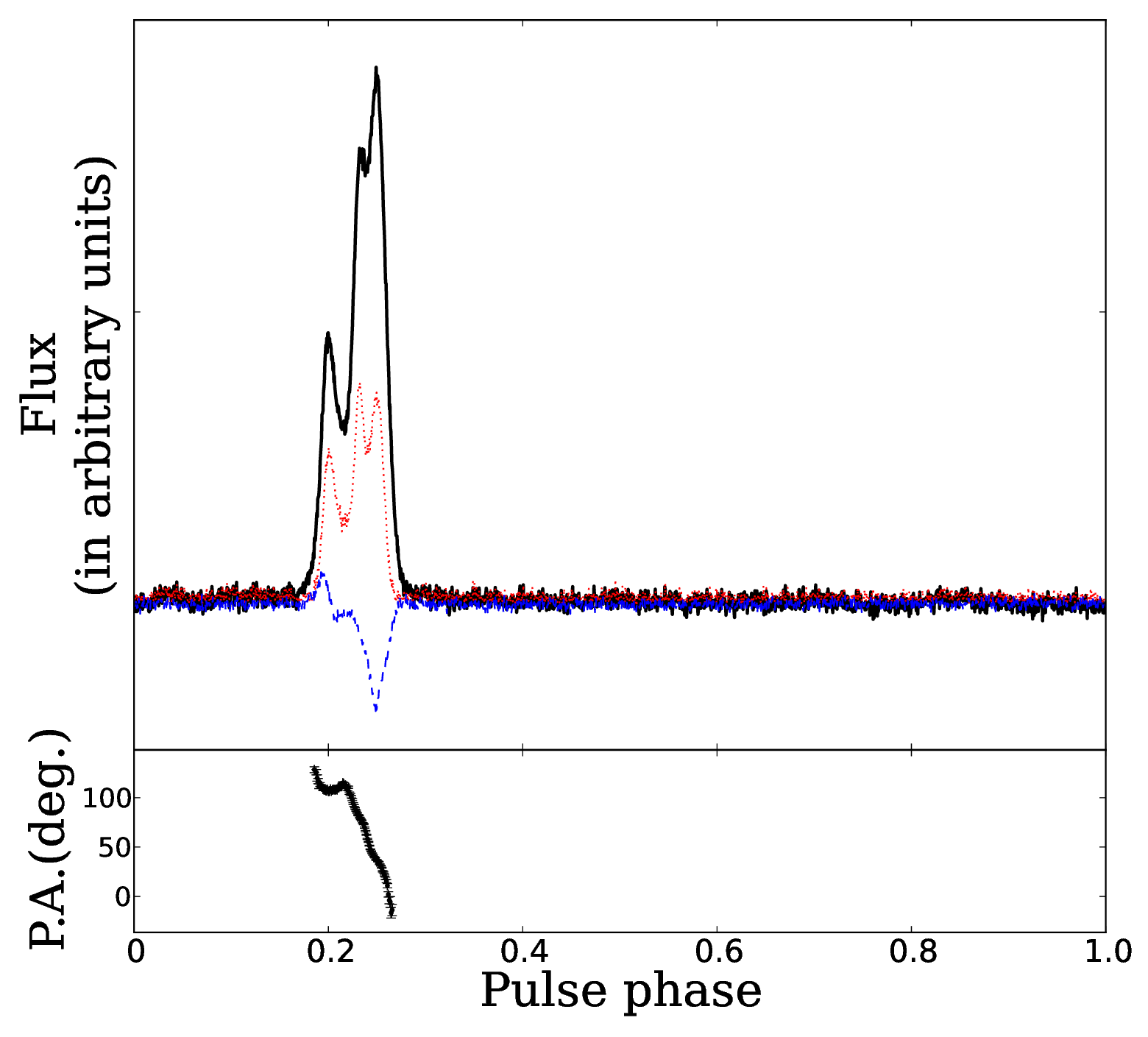}

  \subcaption{}
\label{figure:profile_2045}
\end{subfigure}%
\begin{subfigure}{.5\textwidth}
  \centering
\includegraphics[height = 7.1 cm, angle=0, scale = 1.07]{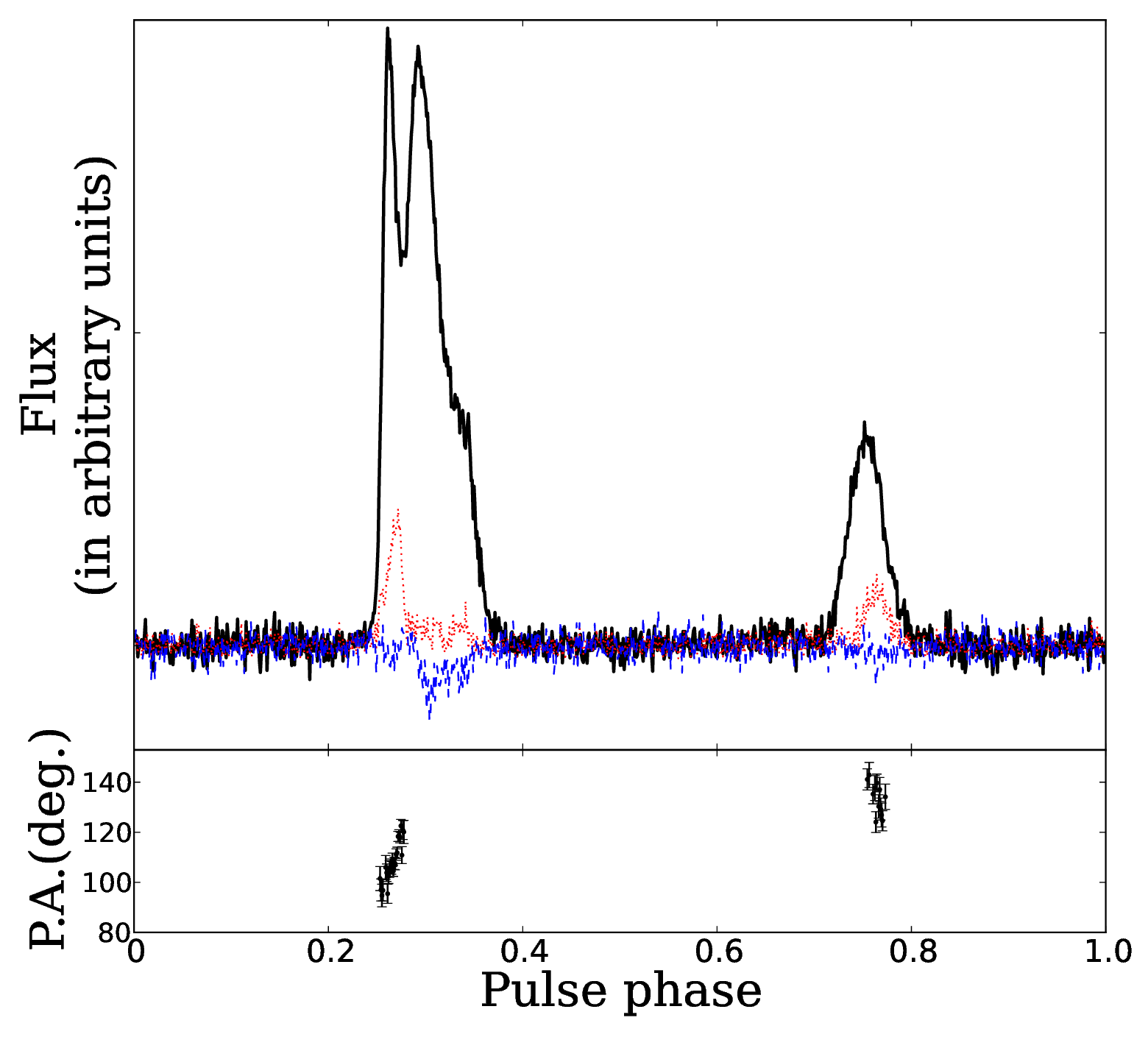}

  \subcaption{}
\label{figure:profile_2053}
\end{subfigure}
\caption{\small \textit{Top} - Average pulse profiles with total intensity (black solid), linear (red dotted)  and circular (blue dashed) polarisation vs. pulse phase for (a) PSR J2045+3633 obtained with the Arecibo telescope at 1430.8 MHz and
(b) PSR J2053+4650 obtained with the Effelsberg telescope at 1347.5 MHz. \textit{Bottom} - Position angle of linear polarisation vs. pulse phase. }
\label{fig:fig2045_2053}
\end{figure*}

\subsection{Discoveries, initial timing and first scientific goals}
 \label{sect:discovery}
 
 Each of the two new bright binary MSPs, PSR J2045+3633 and PSR J2053+4650 (see Fig.~\ref{figure:profile_2045},~\ref{figure:profile_2053}), with spin periods 31.68\, ms and 12.58\, ms, respectively,  was found in 3-minute integration filterbank data 
 down-sampled by factors of four in time and two in frequency, as is usual for survey processing with the quick-look pipeline. This pipeline  \citep[see][]{bck+13} is intended 
 to discover the brightest sources as soon as possible, almost in real time. For this reason it operates on the low-resolution version of the data.
  This pipeline is based on the Fourier transform routines that are typical in pulsar searching.   Before searching for periodic candidates, 
 it performs dedispersion with 406 dispersion measure (DM) trials in the range of 0-2975 pc cm$^{-3}$ in order to correct for  
 {\textit{a priori}} unknown dispersive smearing caused by the propagation of the possible pulsar signal through the ionized interstellar medium. 
 This procedure is sufficient to find most solitary pulsars. In case of pulsars in binary systems whose signals can be modulated by orbital acceleration, 
 to recover the original signal, an additional step of trying many possible acceleration values  may be necessary.
 This is especially crucial for highly relativistic systems in short orbits.  As this task is time-consuming and computationally intensive, the quick-look pipeline does not perform it, 
 thus, limiting our detectability of highly accelerated systems but it still remains sensitive to non-relativistic pulsar binaries with relatively high DMs, like  PSR J2045+3633 and PSR J2053+4650 (129.5 pc cm$^{-3}$
 and 98.08 pc cm$^{-3}$, respectively). 

Confirmation observations for both pulsars were performed with the Effelsberg telescope soon after the discovery. Data taken in search and baseband modes 
for five minutes on several occasions within two weeks after the confirmation showed Doppler-shift variations of the spin period caused by the orbital motion.  
The fitorbit\footnote{https://github.com/gdesvignes/fitorbit/} program was used to produce robust initial parameter estimations prior to commencement of pulsar timing observations.  
Having the ephemeris allowed us to carry out observations in coherent-dedispersion real-time folding mode provided by the PSRIX \citep{lazarus16} backend.
From these data the times-of-arrivals (TOAs) were produced with \psrchive\footnote{http://psrchive.sourceforge.net/} tools for pulsar analysis 
by doing standard procedures of cross-correlating the corresponding profile (obtained by fully integrating the archive file in time and frequency) with a template generated with \texttt{paas}. The initial timing model was fit to the data
 with \tempo\footnote{http://sourceforge.net/} software resulting in a phase-coherent timing solution.

These first results showed that both pulsars are in binary orbits (with periods of 32.3 days for PSR J2045+3633 and 2.45 days for PSR J2053+4650) with massive companions.
Moreover, both systems looked promising for precise measurements of the masses of the components through determination of post-Keplerian parameters \citep{TW1982}. PSR J2045+3633 has revealed an 
eccentricity $e\, = \, 0.0172 \ $, making possible a measurement of the rate of  periastron advance.
PSR J2053+4650 seemed to be in a circular, but highly-inclined  orbit allowing the possible detection of the Shapiro delay \citep{Shapiro64} 
caused by the propagation of the pulsar signal in the gravitational field of the companion
(the effect is most noticeable when the orbit is close to being visible edge-on, i.e.\ the orbital inclination angle $i \approx 90^{\circ}$). The low root-mean-square (rms)
 of the first timing residuals (30 to 40 $\mu$s for each pulsar) indicated that 
significant measurements of these parameters could be made on a short-term basis (within a few years).

\section{High precision timing and analysis}
 \label{sect:timing}

\centering
\begin{table*}  
\centering
\caption{Timing observations of PSRs J2045+3633 and J2053+4650 with four telescopes}
\begin{tabular}{l c c c c }
\hline\hline
Telescope & Effelsberg  & Jodrell Bank & Nançay  & Arecibo\\
\hline
Backend   & PSRIX  &  ROACH    & NUPPI   & PUPPI      \\
Centre frequency  (MHz) & 1347 & 1520 & 1484 & 1430 \\
Effective bandwidth (MHz) & 240 & 384 & 512 & 600 \\
Integration time (min) & 30 & 30 & 10 & 11 \\
\hline
\multicolumn{5}{c}{Observation Parameters for PSR J2045+3633}\\ 
\hline
Number of TOAs    & 274 & 75 & 214 & 30 \\ 
Number of frequency subbands (for producing TOAs)    & 4 & 1 & 4$^{*}$ & 1 \\ 
Weighted RMS of post-fit Timing Residuals ($\mu$s) & 5.388&  4.356 & 7.541   & 1.110 \\
EFAC &  1.297   &   1.033   &   1.245   &  2.757 \\
Date span (MJD)  & 56996-57284 & 56911-57538  & 57097-57506 & 57258-57294 \\
\hline
\multicolumn{5}{c}{Observation Parameters for PSR J2053+4650} \\
\hline
Number of TOAs   &  121   &   79      &   600     &  ---    \\
Number of frequency subbands (for producing TOAs)    & 2 & 1 & 4$^{*}$ & --- \\ 
Weighted RMS of post-fit Timing Residuals ($\mu$s) &   3.078   &  3.128  & 5.189   & ---  \\
EFAC &  1.318    &  1.208   &    1.145   & --- \\
Date span (MJD)   & 56996-57145 & 56911-57538  & 57097-57490 & ---  \\
\hline
\end{tabular}
\begin{flushleft}
$^{*}$ In most cases we kept four subbands but for some epochs the TOAs with uncertainties larger than 8 $\mu$s were excluded from the analysis resulting in two or three TOAs per epoch.\\
\end{flushleft}
\label{table:telescopes} 
\end{table*}

\justifying

Regular timing  was started in September 2014. The general timing strategy (i.e., excluding special campaigns - see~\ref{Sect:Sp2045} and~\ref{Sect:Sp2053}) for both pulsars was similar: they were systematically followed up with the Lovell, Effelsberg and
\textendash\ (since March 2015)  \textendash\ Nan\c{c}ay radio telescopes, with an  observation lasting, on average, 30-60 minutes. At Jodrell Bank, the pulsars were observed almost daily during the first three weeks
and later - once in 10-20 days, at Effelsberg - almost monthly since December 2014 until November 2015, at Nan\c{c}ay  the cadence varied from every day to once in a few months.
Observational parameters and the details of the recording systems are presented in Table~\ref{table:telescopes}. 

The overall  preparation procedure was the same for the data from all the observatories: using \psrchive\ tools, the data from each telescope were cleaned of RFI, fully integrated in 
time  and integrated in frequency to keep a number of subbands from one to four (depending on the observatory - see Table~\ref{table:telescopes}) 
and  cross-correlated with a telescope-specific template to create TOAs. 
For the Lovell and Effelsberg observations we used an analytic template fit to the high signal-to-noise data created with the \texttt{paas} routine from \psrchive. For Nançay  
the template was produced by adding 10 high signal-to-noise profiles for each pulsar, 
and the results from these additions were then smoothed. For the highest-quality Arecibo  data the template was constructed from the average Arecibo pulse profiles over observed epochs.
In our timing analysis we used \tempotwo\ software package \citep{hem06} refining the timing model by least-square fitting 
the parameters of the system. The TOAs from different observatories were converted to the Solar System barycentre using the DE421 ephemeris. The details of fitting and models used are described in Section~\ref{sec:masses}.

Given the eccentricity of PSR J2045+3633 and the fortuitous orbital inclination of PSR J2053+4650, we initiated special timing campaigns  aiming to improve  the measurement of the rate of advance of periastron 
and, possibly, the Shapiro delay for PSR J2045+3633, and the Shapiro delay for PSR J2053+4650, with a goal of constraining the masses of the pulsar and  companion in both systems.

\subsection{Special campaigns: PSR J2045+3633}
\label{Sect:Sp2045}
For PSR J2045+3633 we conducted two full-orbit campaigns: one with the Effelsberg telescope and, in order to maximise our sensitivity, one with Arecibo.

The 32-day Effelsberg  campaign took place in April-May 2015 and consisted of 10 four-hour observations at a central frequency of 1.4 GHz spread across the orbit.
The data were recorded   with  the PSRIX  \citep{lazarus16} backend  in   its coherent-dedispersion real-time folding mode (the same system that is used for regular timing).
The average TOA error was 3 $\mu$s.

\begin{figure}
 \includegraphics[width=5.75in, angle=0,scale=0.54]{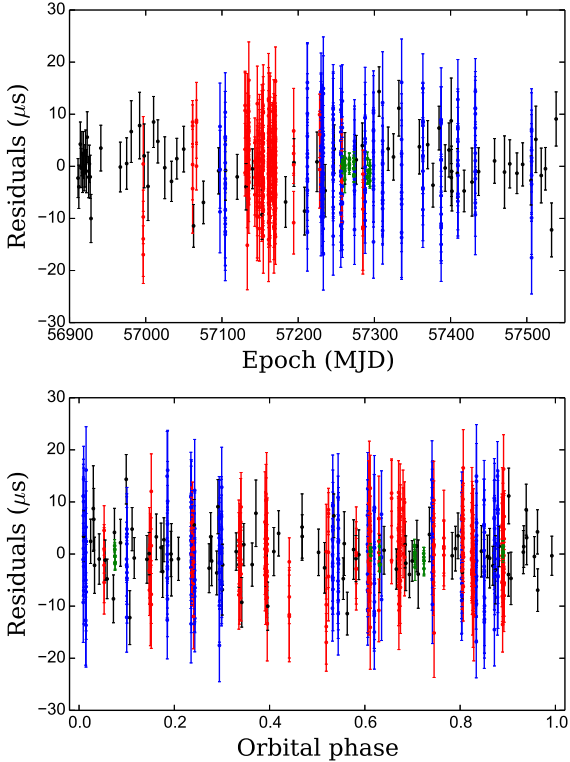}
 \caption{\small The top plot shows the post-fit timing residuals for PSR J2045+3633 as a function of MJD,  the bottom plot - as a function of orbital phase.  The residuals obtained from the data taken with different telescopes are 
marked with different colours: Jodrell Bank - black,  Nançay - blue, Effelsberg - red, Arecibo - green. The error bars represent the 1-$\sigma$ uncertainties of TOA measurements. }
\label{figure:2045_TOAs}
\end{figure}

The Arecibo campaign was held in August-September 2015 with the use of the L-band wide receiver. The data were coherently folded in real time with  the PUPPI (Puerto Rico Ultimate Pulsar Processing Instrument) 
backend (Table~\ref{table:telescopes}). We performed 6 full-transit (55-minute) observations covering a part of the orbit. 
Due to some issues with the receiver, it was not possible to cover the whole orbit as planned but the data obtained (combined with all the other TOAs) still allowed us to measure the rate of advance of
periastron ($\dot{\omega}$) at 5-${\sigma}$ level: 0.0010(2)$^\circ \rm yr^{-1}$ - and an indication of the Shapiro delay with 3- and 2-${\sigma}$ measurement of its orthometric amplitude and ratio:  $h_3= 1.0(3)$ $\mu$s  
and $\varsigma = 0.6(3)$ respectively (see Section~\ref{sec:masses}).

The average TOA error from this campaign was 0.29 $\mu$s for the 11-minute observations fully scrunched in time and frequency.
The higher precision of these TOAs compared to the ones from other telescopes made them overweighted in the overall timing analysis. 
For this reason, TOA uncertainties for a set of data from every observatory were multiplied by a factor (EFAC) between 1.033 and 2.757 
(see Table~\ref{table:telescopes})
to achieve reduced $\chi^{2}=1$.  EFACs of Jodrell Bank, Effelsberg and Nan\c{c}ay TOAs, slightly differing from unity,  can be explained by  the presence of RFI in the data whereas a much larger
EFAC of much more precise Arecibo TOAs is probably an evidence of the intrinsic pulsar red noise. The overall timing residuals as a function of MJD and orbital phase are presented in Fig.~\ref{figure:2045_TOAs}.

\begin{table*}
  \centering
  \caption{Timing parameters for PSRs J2045+3633 and J2053+4650}
  \label{table:parameters}
   \begin{tabular}{l c c c}
  \hline
  \hline
   Pulsar name & PSR J2045+3633 & \multicolumn{2}{c}{PSR J2053+4650} \\
   Binary model & DDH & DDH & DD \\
   Solar System ephemeris & DE421 & DE421 & DE421 \\
  \hline
  \multicolumn{4}{c}{Spin and astrometric parameters}\\
  \hline
  Right ascension, $\alpha$ (J2000)  &  20:45:01.50504(12) & \multicolumn{2}{c}{20:53:52.62804(7)} \\
  Declination, $\delta$ (J2000)  &  +36:33:01.4033(8) &   \multicolumn{2}{c}{+46:50:51.7181(4)}    \\
  Proper motion in RA, $\mu_{\alpha}$ (mas yr$^{-1}$)  &  $-$2.1(1.4) &   \multicolumn{2}{c}{$-$2.8(8)}  \\
  Proper motion in DEC, $\mu_{\delta}$ (mas yr$^{-1}$) &  $-$2.3(8)  &   \multicolumn{2}{c}{$-$5.4(5)}  \\
  Total proper motion, $\mu_{\rm{tot}}$ (mas yr$^{-1}$)  &  3.1(1.1) &  \multicolumn{2}{c}{6.1(5)} \\
  Galactic longitude, $l$  & 77.83 &  \multicolumn{2}{c}{86.86}  \\
  Galactic latitude, $b$  & $-$3.93  &  \multicolumn{2}{c}{1.30}   \\
  Pulse frequency, $\nu$ (s$^{-1}$) & 31.56382007686(1) &  \multicolumn{2}{c}{79.45162290069(1)}  \\ 
  First derivative of pulse frequency, $\dot{\nu}$ (s$^{-2}$) & $-$5.861(4)$\times 10^{-16}$ &   \multicolumn{2}{c}{1.0875(6)$\times 10^{-15}$} \\ 
  Spin period  $P$ (ms)  & 31.68184324854(1) & \multicolumn{2}{c}{12.586275314350(2)} \\
  Observed period derivative, $\dot{P}$ ($10^{-19}$ s s$^{-1}$)  &  5.883(3) &   \multicolumn{2}{c}{1.7229(8)} \\
  Dispersion measure, DM (pc cm$^{-3}$)  &  129.5477(17) &   \multicolumn{2}{c}{98.0828(6)}  \\
  Rotation measure, (rad m$^{-2}$) & $-$266(10) & \multicolumn{2}{c}{$-$174(11)} \\ 
  \hline
  \multicolumn{4}{c}{Binary Parameters}\\
  \hline
  Orbital period, $P_b$ (days)  &  32.297845(1)  &  \multicolumn{2}{c}{2.4524990114(2)}  \\
  Projected semi-major axis of the pulsar orbit, $x$ (lt-s) & 46.941885(11) &  8.8042995(11) &  8.8042996(11) \\
  Epoch of periastron, $T_0$ (MJD)  &  57496.75108(3) & \multicolumn{2}{c}{56911.113(3)} \\
  Orbital eccentricity, $e$  &   0.01721244(5)  & \multicolumn{2}{c}{0.0000089(1)}    \\
  Longitude of periastron, $\omega$ ($^\circ$)  &  320.7822(3) & \multicolumn{2}{c}{266.7(4)} \\
  \hline
  \multicolumn{4}{c}{Relativistic parameters and masses} \\
  \hline
  Rate of advance of periastron, $\dot{\omega}$ ($^\circ \rm yr^{-1}$)  &  0.00105(14) & --- & --- \\
  Orthometric amplitude, $h_3$ ($\mu$s)  &  1.0(3) &3.23(15) & --- \\
  Orthometric ratio, $\varsigma$  & 0.6(3) &  0.918(14) & ---  \\
  Orbital inclination, $i$ ($^\circ$)   & $62\substack{+ 5 \\ -6}$ &  $85.0\substack{+ 0.8 \\ -0.9}$ &  $85.1\substack{+ 0.9 \\ -0.7}$  \\
  \hline
  Mass function, $f$ (\Msun)  & 0.1064621(2)  &  \multicolumn{2}{c}{0.12182741(4)}   \\
  Total mass, $M$ (\Msun)  & 2.28(45) (derived from $\dot{\omega}$ )  & ---  & 2.23(24)$^{a}$ \\
  Pulsar mass, $m_{p}$ (\Msun)  & $1.33\substack{+ 0.30 \\ -0.28}^{b}$ & $1.40\substack{+ 0.21 \\ -0.18}^{b}$  & 1.38(18)$^{a}$ \\
  Companion mass , $m_{c}$ (\Msun)  & $0.94\substack{+ 0.14 \\ -0.13}^{b}$ & $0.86\substack{+ 0.07 \\ -0.06}^{b}$ & 0.85(6)$^{a}$ \\
  \hline
  \multicolumn{4}{c}{Derived parameters}\\
  \hline 
  DM-derived distance (NE2001)$^{c}$, $d$ (kpc)  & 5.51 &   \multicolumn{2}{c}{4.12} \\   
  DM-derived distance (YMW16)$^{d}$, $d$ (kpc)  & 5.63 &   \multicolumn{2}{c}{3.81}\\
  Shklovskii's correction to period derivative, $\dot{P}$ ($10^{-21}$ s s$^{-1}$) & 4.2(3.1) & \multicolumn{2}{c}{6.4(2.2)}\\
  Shklovskii-corrected period derivative, $\dot{P}$ ($10^{-19}$ s s$^{-1}$)  & 5.84(3)  & \multicolumn{2}{c}{1.66(2)}  \\
  Surface magnetic field strength, $B_0$ ($10^{9}$ Gauss)  & 4.1  &  \multicolumn{2}{c}{1.4} \\
  Characteristic age, $\tau_c$ (Gyr)  & 0.85 &  \multicolumn{2}{c}{1.15}  \\
  \hline
\end{tabular} 

\begin{flushleft}
$^{a}$ The pulsar and companion masses are calculated assuming GR from the range and shape of the Shapiro delay. The total mass is then calculated by summing these values.\\
$^{b}$ The pulsar and companion masses derived from the Bayesian mapping  (see sections about mass measurements). \\
$^{c}$ We assume the uncertainty on the distance to be 25-30 \%, what is commonly accepted when using the \nemodel\  model \citep{cl02}, though these numbers represent a very average estimate 
and in general depend on the line of sight.  \\
$^{d}$ We assume the uncertainty on the distance to be 20-40 \%, though it may be significantly underestimated \citep{2017YMW}.\\
\footnotesize {Figures in parentheses are  the nominal 1-$\sigma$ \textsc{tempo2} uncertainties in the least-significant digits quoted.}\\
\end{flushleft}
\end{table*}

\subsection{Special campaign: PSR J2053+4650}
\label{Sect:Sp2053}

\begin{figure}
 \includegraphics[width=5.75in, angle=0,scale=0.54]{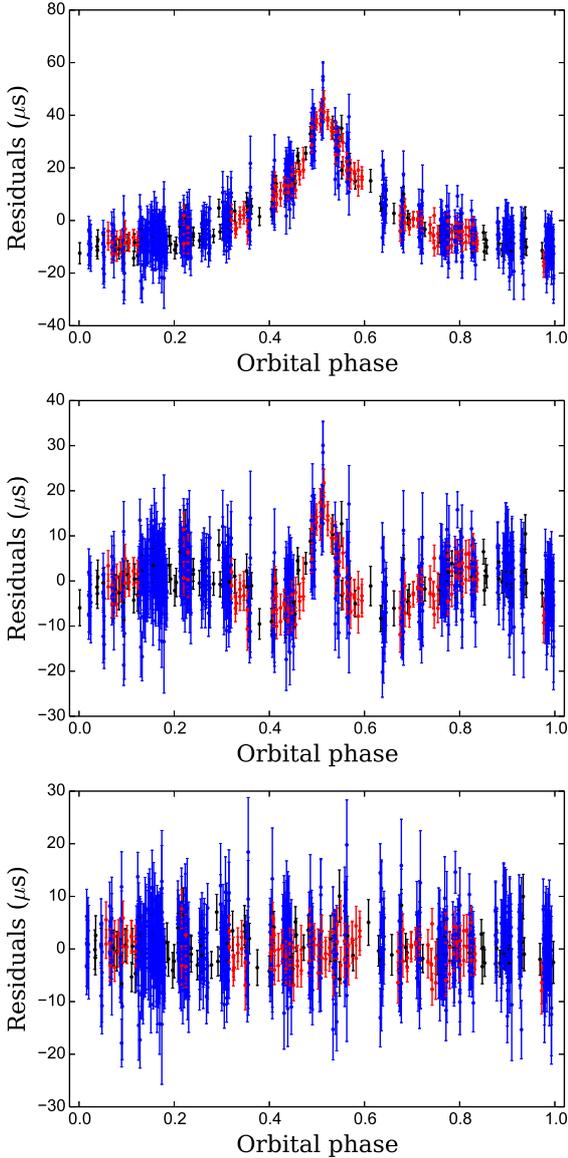}
 \caption{Timing residuals for PSR J2053+4650 as a function of orbital phase: the top plot shows residuals before fitting for Shapiro delay parameters, the bottom plot - after fitting. 
 The middle plot shows timing residuals after fitting for Keplerian parameters: a part of the Shapiro delay is absorbed by this fit. The residuals obtained from the data taken with different telescopes are 
marked with different colours: Jodrell Bank - black,  Nançay - blue, Effelsberg - red.  The error bars represent the 1-$\sigma$ uncertainties of TOA measurements.}
\label{fig:resids_2053}
\end{figure}

For PSR J2053+4650 we performed two successive full-orbit campaigns with the Effelsberg telescope in the course of five days (April 29th-May 3rd 2015).  On the first four days we observed it for four hours every day,
and on the fifth day  we took a twelve-hour session near the superior conjunction. This campaign helped to make a very significant Shapiro delay detection.  Combined with the other TOAs (see Fig.~\ref{fig:resids_2053}), 
this resulted in the  22- and 66-$\sigma$ measurement of the orthometric amplitude and orthometric ratio respectively:  $h_3= 3.23(15)$ $\mu$s  and $\varsigma = 0.918(14)$  (see Section~\ref{sec:masses}). 

\section{Results and discussion}

The resulting best-fit timing parameters for both pulsars are presented in Table~\ref{table:parameters}. These include spin, astrometric and orbital parameters,
as well as the derived masses of the components of the systems.

\subsection{Polarisation studies}
\label{Sect:pol}

\begin{figure*}
\includegraphics[width=6in, scale = 0.4]{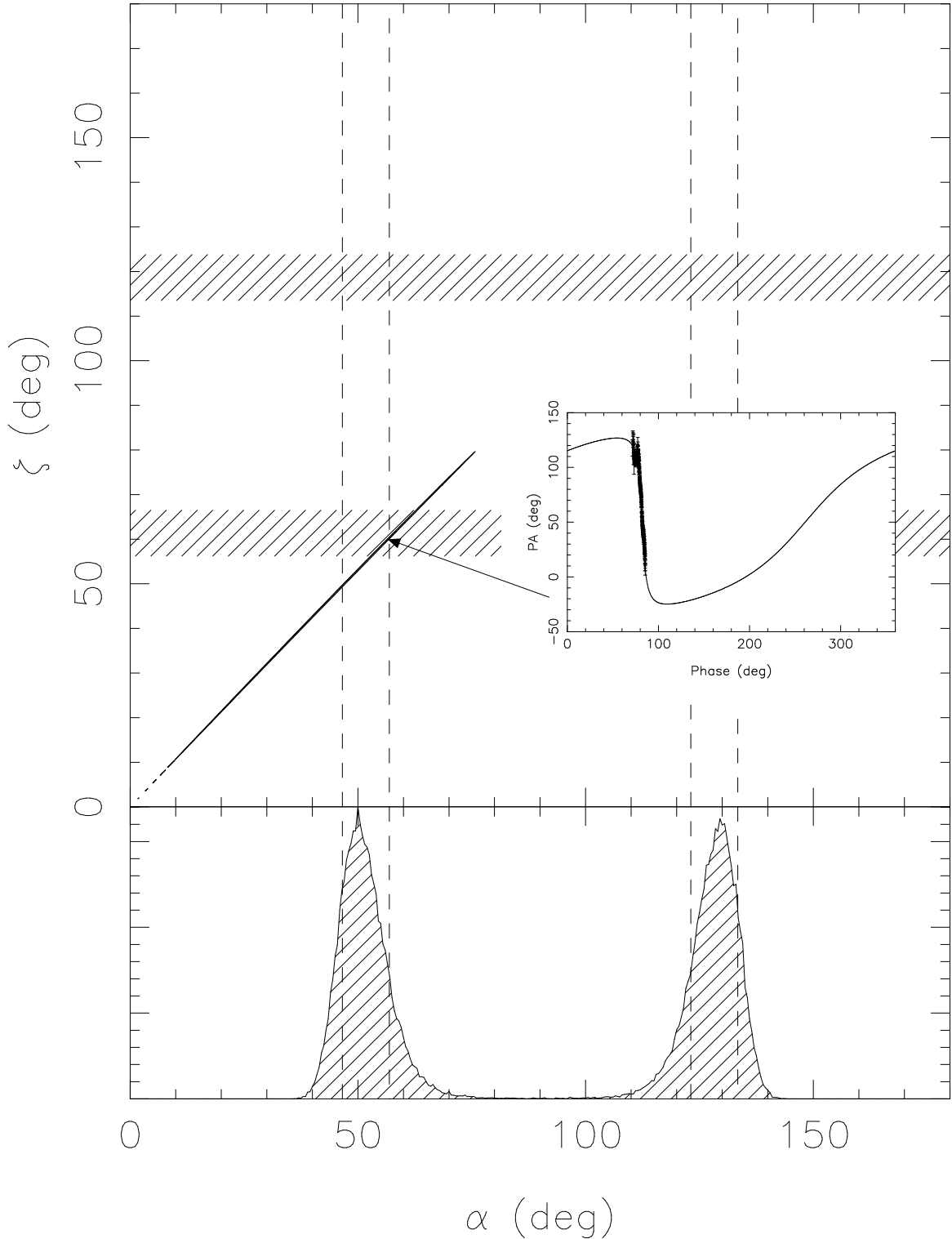}
\caption{System geometry for 
PSR J2045+3633 as derived from a least-squares-fit of
the Rotating Vector Model (RVM) to
the position angle (PA) of the linearly polarized emission.
The top panel shows the $\chi^2$ contours from minimizing
$\chi^2$ by stepping through values of the  magnetic inclination angle ($\alpha$) and
viewing angle $\zeta$ while simultaneously minimizing reference phase
$\Phi_0$ and reference position angle $\Psi_0$ of the RVM at each grid point.
In order to constrain the geometry further in the presence of the
correlation between $\alpha$ and $\zeta$, we mark the constraints
on the orbital inclination angle as horizontal strips (see text for
details). Also, assuming a filled emission beam, we derive a
distribution of magnetic inclination angles (lower panel) that is consistent
with the observed pulse width (see text for details). The vertical
dashed lines indicate a $\pm34$\% range around the median
value of the two preferred solutions. The solutions for smaller $\alpha$ are more
consistent with the joint constraints from the RVM fit and radio
timing, indicating that the true underlying orbital inclination angle
is $i\sim60^{\circ}$ (rather than
$180^{\circ}-60^{\circ}=120^{\circ}$). A corresponding RVM fit for
$\alpha=57^{\circ}$ and $\zeta=60^{\circ}$ is shown in the small
 insert in the middle of the figure.}
\label{fig:cont}
\end{figure*}
 
At Effelsberg and Arecibo we performed a polarisation calibration
observation with the noise diode prior to every pulsar observation.
The diode signal was injected into the receiver feed horn at an angle
of $45^{\circ}$ to both polarisation probes when the telescope was
$0.5^{\circ}$ off source.  These observations were used to calibrate
the pulsar data with \texttt{pac} from \psrchive.  Several calibrated
observations were then integrated to obtain a low-noise polarisation
profile (Fig.~\ref{figure:profile_2045},~\ref{figure:profile_2053}:
for PSR J2045+3633 we present the Arecibo profile as it has much
higher signal-to-noise ratio than the one from Effelsberg).
The data for both pulsars  were corrected for Faraday rotation with the values of rotation measure
determined using  \texttt{rmfit} from \psrchive\ (see Table~\ref{table:parameters}). 

PSR J2045+3633 shows a significant degree of both linear and circular
polarisation (Fig.~\ref{figure:profile_2045}).  Interpreted in the
framework of polarisation features found in non-recycled pulsars
\citep[e.g.][]{handbook}, the change from a weak positive circular
polarisation to significant negative polarisation suggests a central
line of sight through the beam. This is consistent with the observed
steep position angle (PA) swing. Despite the flat PA values in the
leading part of the PA swing, it can be well described by a standard
rotating vector model (RVM) fit, which in principle allows us to
constrain the viewing geometry \citep[e.g.][]{handbook}. The resulting
$\chi^2$ values from a least-squares fit of the RVM to the PA swing
produce the contours in the magnetic inclination angle ($\alpha$) and
viewing angle (i.e.\ angle between the spin axis and the observer’s
direction, $\zeta$) plane shown in Fig.~\ref{fig:cont}. 

The correlation between $\alpha$ and $\zeta$ is not surprising and is well
known; note that $\zeta=\alpha+\beta$, where $\beta$ is the impact
angle, i.e.\ the angle between magnetic axis and observer at the
closest approach. In order to constrain the geometry further, we can
utilize the constraints on the orbital inclination angle from pulsar
timing, since for a pulsar beam of angular radius $\rho$ to be visible
to the observer, we find $|i-\zeta |<\sim\rho$. The constraints on $i$
and $180^{\circ}-i$ are correspondingly plotted in the upper panel of
Fig.~\ref{fig:cont} as horizontal strips with a width identical to
their uncertainties. For a pulsar with a filled emission beam, the
observed pulse width $w$ is a function of $\alpha$, $\zeta$ and $\rho$
\citep[e.g.][]{handbook}. For non-recycled pulsars, we find
$\rho = k /\sqrt{P}$, where $P$ is the pulse period and value of $k$
depending weakly on frequency. The exact value differs between
different authors, but for a width measured at a 10\% intensity level
$k$ is typically about $6.3^{\circ}$ \citep{kramer1994}.  Adopting
this relationship and measuring a width of $w=36^{\circ}$ we performed
Monte Carlo simulations, where we drew $\rho$ from a normal
distribution centred on $k=6.3^{\circ}$ with a width of $0.63^{\circ}$
as a typical uncertainty reflecting both the uncertainty in $k$ and
the measured width. We also drew $\alpha$ from a flat distribution
between 0 and 180$^{\circ}$, while using another flat distribution for
$\beta\le \rho$ (i.e.\ the condition for the observer to register the
beam). Testing 10 million combinations, we recorded those values of
$\alpha$ where the observed width was consistent with the simulated
values.  This distribution of $\alpha$ is shown in the bottom panel of
Fig.~\ref{fig:cont}.
Two preferred ranges of solutions near
50$^{\circ}$ and 130$^{\circ}$ are clearly visible. Dashed vertical
lines indicate a $\pm34$\% range around the median
value. Interestingly, the solutions for smaller $\alpha$ are more
consistent with the joint constraints from the RVM fit and radio
timing, indicating that the true underlying orbital inclination angle
is $i\sim60^{\circ}$ (rather than
$180^{\circ}-60^{\circ}=120^{\circ}$). A corresponding RVM fit for
$\alpha=57^{\circ}$ and $\zeta=60^{\circ}$ is shown in the small
insert in the middle of the figure. In summary, radio timing, PA swing, and profile
width data can all be explained in a self-consistent geometric model,
making assumptions that were derived from normal (non-recycled)
pulsars.

\begin{figure*}
\includegraphics[width=6in, scale = 0.4]{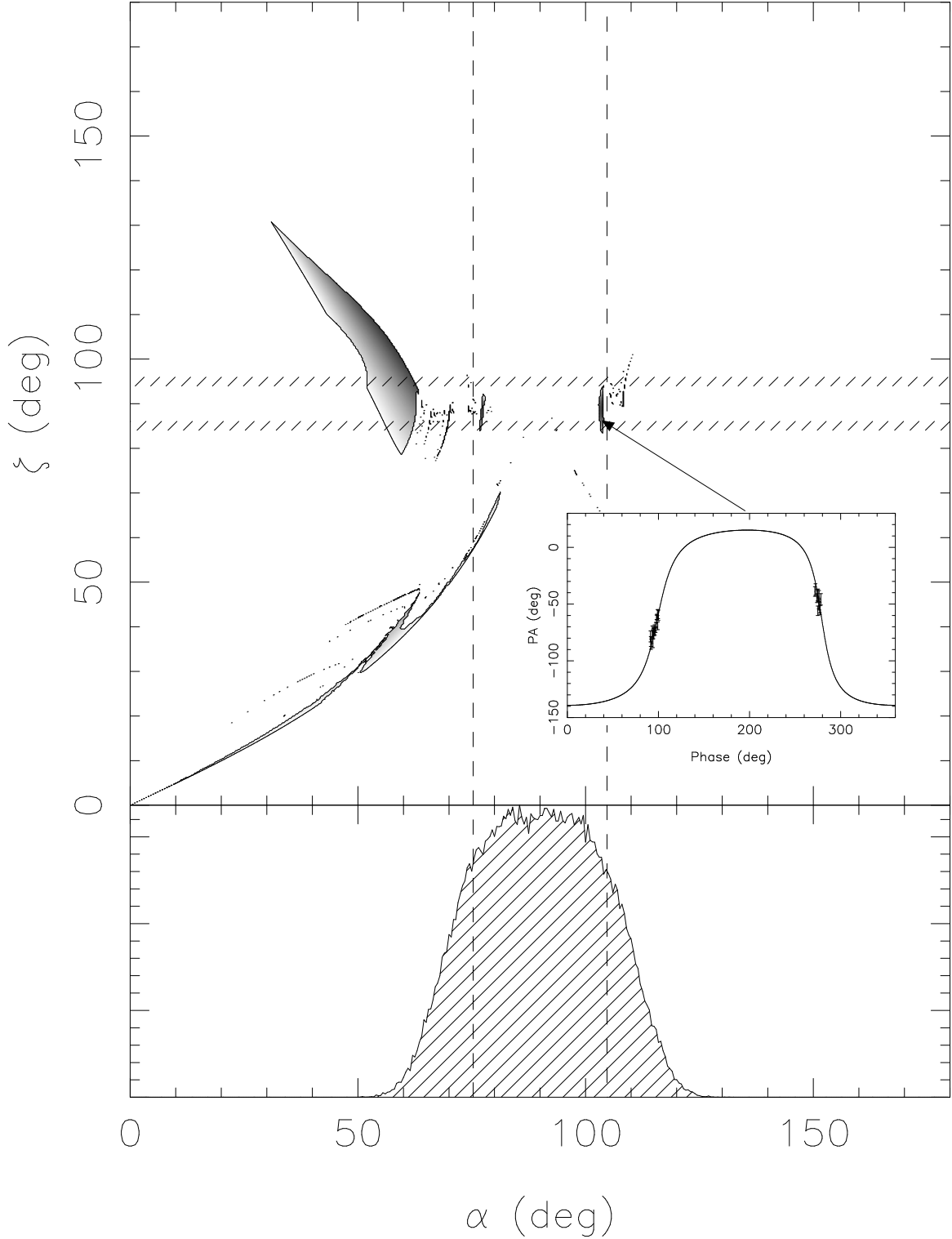}
\caption{System geometry for 
PSR J2053+4650 as derived from least-squares-fit of
the Rotating Vector Model to
the position angle  of the linearly polarized emission.
See Figure~\ref{fig:cont} and text for details.}
\label{fig:cont2053}
\end{figure*}

We repeat the same procedure of fitting the RVM model for PSR
J2053+4650. We can identify two main groups of significant PA values,
clustering around longitudes of $100^{\circ}$ and $280^{\circ}$, respectively (see insert in Fig.~\ref{fig:cont2053}).
We applied the same method as described above, but allow for the
possibility that the two main clusters of PA values are separated by
phase offsets of $\pm90^{\circ}$ and $\pm180^{\circ}$. This leads to
various ``islands'' in the shown $\chi^2$ plot. The indicated best fit
solution that is consistent with a measurement of the orbital
inclination angle (Table~\ref{table:parameters}) is derived for zero
offset between the two PA clusters. If we ignore the big ``island''
outside the alpha range, then the smaller inclination angle
($87.5^\circ$) is slightly preferred but it is much less clear than in
the case of PSR J2045+3633. What is clear from the given constraints
is that the best solution is that of an orthogonal rotator, which is
also consistent with the observation of an interpulse. Given this
nearly orthogonal viewing geometry of the pulsar, it is to be expected
that both solutions for the inclination should indeed have nearly
equal probability.

\subsection{Astrometric parameters}

High-precision timing observations  conducted for both binaries for almost 21 months gave us an opportunity to measure their proper motions.  
As can be seen from Table~\ref{table:parameters}, for PSR J2045+3633 the proper motion is poorly constrained at $\mu_{\rm{tot}} = 3.1 \pm 1.1\, \rm mas\, yr^{-1}$.  
 For PSR J2053+4650 the significance is higher at $\mu_{\rm{tot}}\, =\, 6.1\, \pm\, 0.5\, \rm mas \, yr^{-1}$. Knowing the total proper motion, we can derive the transverse velocity from:
\begin{equation}
\frac{v_{\rm t}}{\rm km\,s^{-1}} = \rm4.74\times \left(\frac{\mu_{tot}}{\rm mas\,yr^{-1}}\right) \times \left(\frac{d}{\rm kpc}\right),
\end{equation}
where d is the distance to the pulsar. According to the  \nemodel\  model described in \cite{cl02}, assuming the 25-30 \% uncertainty on the DM-derived distance 
 \citep[although it should be noted that these uncertainties may be underestimated, see][]{2009Deller},  we obtain 
$v_{\rm t}\, =\, 80 \, \pm 40\, \rm  km \, s^{-1}$ 
for PSR J2045+3633 and  $v_{\rm t}\, =\, 120 \pm 40\, \rm km \, s^{-1}$ for PSR J2053+4650. Using the newer YMW20016 model \citep{2017YMW} and assuming an average distance error of 40 \% (though for individual sources
it may be significantly larger), we get  
$v_{\rm t}\, =\, 82 \, \pm 40\, \rm  km \, s^{-1}$ for PSR J2045+3633 and  $v_{\rm t}\, =\, 110 \pm 50\, \rm km \, s^{-1}$ for PSR J2053+4650. 
The values obtained within the both models are consistent with each other and with the transverse velocities observed for the general population of binary MSPs \citep{hobbs2005,gon11,desvignes16}.

As demonstrated by  \citet{shk70}, the apparent period derivative of a pulsar is affected by its transverse motion as:
\begin{equation}
 \frac{\dot{P}}{P}\, =\, \frac{1}{c}\,\times \frac{{v_{\rm{t}}^2}}{d}\,,
\end{equation}
where $c$ is the speed of light.
As can be seen from Table~\ref{table:parameters}, currently the precision of  the determined Shklovskii contribution is low for both MSPs: almost 1$\sigma$ for PSR J2045+3633 and 3$\sigma$ 
for PSR J2053+4650. However, they are two orders of magnitude below (${\sim}10^{-21}$ s) the observed $\dot{P}$ values (${\sim}10^{-19}$ s), so they do not affect $\dot{P}$ significantly. 
Nevertheless, we used the corrected values for estimating the magnetic field strengths and the characteristic ages.

\subsection{Mass measurements and the nature of the companions}
\label{sec:masses}

\begin{figure*}
\includegraphics[width=7in]{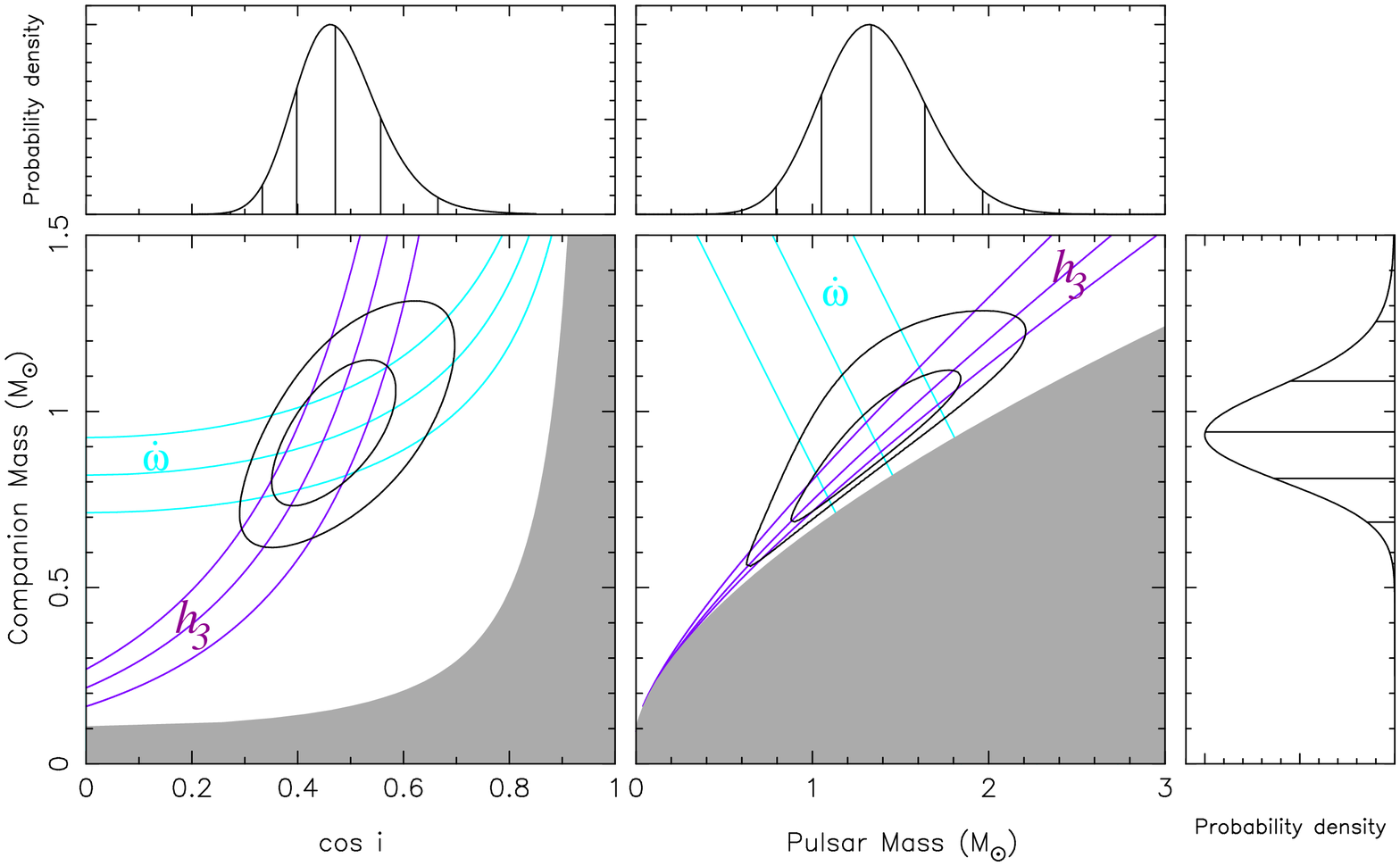}
\caption{\small Constraints on the masses of the components  and the orbital inclination angle for PSR
  J2045+3633. Each triplet of lines corresponds to the median values and
  $\pm 1 \sigma$ uncertainties of the post-Keplerian parameters obtained 
  within the DDH model: the rate of advance of periastron $\dot{\omega}$ (cyan) and the orthometric
  amplitude of the Shapiro delay, $h_3$ (purple) (the triplet for the 
  orthometric ratio of the Shapiro delay $\varsigma$ is removed from the figure  for better visual perception as it covers a wide region which includes both regions for $\dot{\omega}$ and  $h_3$)
  The contour levels contain 68.27 and 95.45\% of the 2-D probability
  density functions (pdfs) derived from the quality of the timing solution
  at each point on the $M_c$ - $\cos i$ plane using the Shapiro
  delay together with an assumption that $\dot{\omega}$ can be fully described by general relativity  to constrain the masses.
  The left plot shows the $M_c$ - $\cos i$ plane with the gray region excluded by
  the physical constraint $M_p > 0$. The right plot shows $M_c$ - $M_p$
  plane with the gray region excluded by the mathematical constraint
  $\sin i \leq 1$.  The top plots depict probability density functions for $\cos i$, $M_p$
  and  the right marginal plot - for $M_c$, derived from
  marginalizing the 2-D pdf in the main panel for these
  quantities.}
\label{figure:2045_mass}
\end{figure*}

\begin{figure*}
\includegraphics[width=7in]{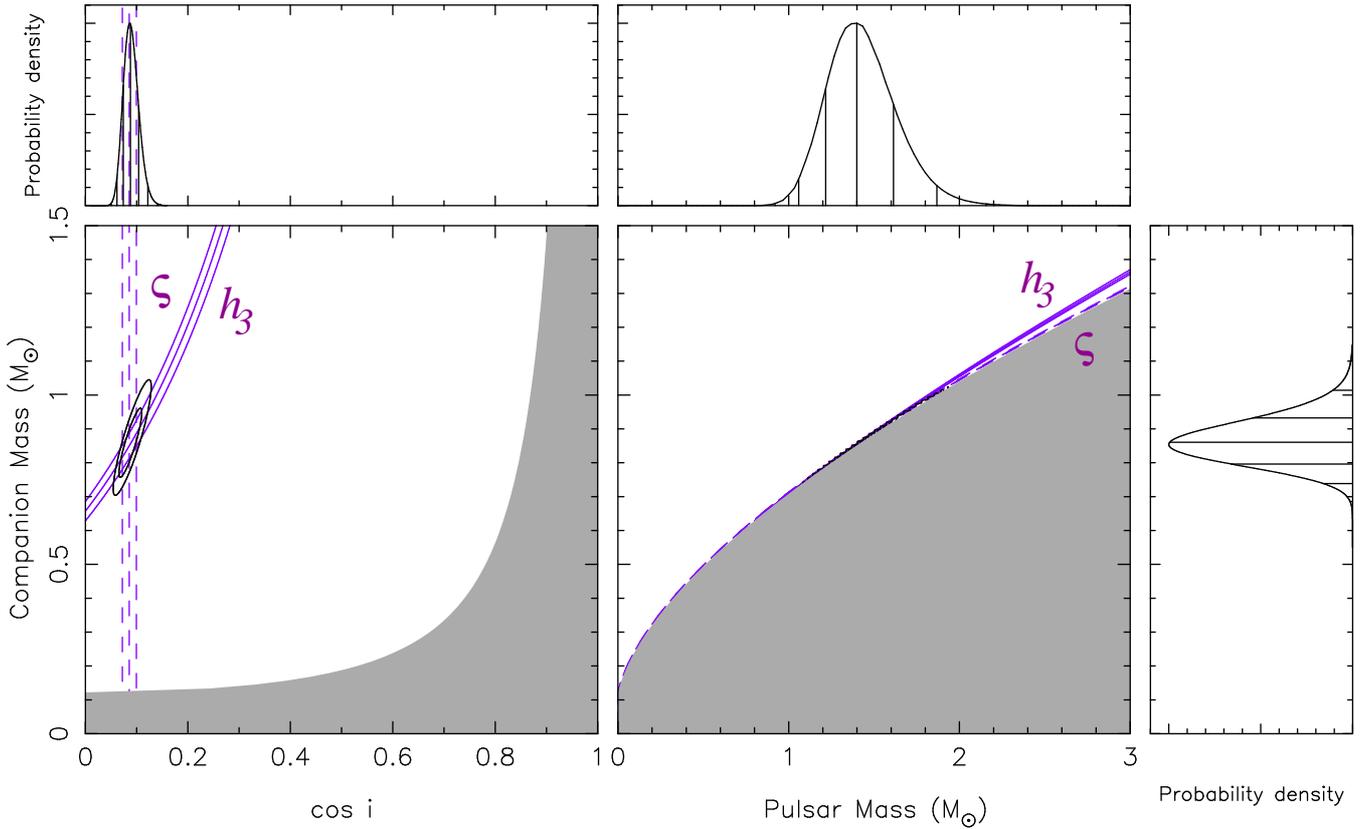}
\caption{\small  Constraints on the masses of the components and the orbital inclination angle for PSR
  J2053+4650 in the same manner as in Fig.~\ref{figure:2045_mass}.
  The only difference is in the PK parameters used to constrain the masses:
  in this case the orthometric amplitude $h_3$ (solid purple)
  and the orthometric ratio of the Shapiro delay $\varsigma$ (dashed purple).}
\label{figure:2053_mass}
\end{figure*}

The masses of the components of the system are related to each other and to the Keplerian parameters of the orbit through the mass function \citep[e.g.][]{handbook}:
\begin{equation} 
\label{eq:mass_function}
f(M_p,M_c) = \frac{(M_{c}\sin{i})^{3}}{(M_{p}+M_{c})^2} = \frac{4\pi^{2}}{T_{\odot}}\frac{x^{3}}{P_b^{2}},
\end{equation}
where $M_p$ and $M_c$ are the pulsar and companion masses in units of
a solar mass, $i$ is the orbital inclination, $x$ is the projected semi-major axis of the orbit in light seconds and
$\rm T_{\odot} \equiv G  M_{\odot} c^{-3} = 4.9254909476412675\, \mu
s$ is a solar mass in time units (in the latter expression
$c$ is the speed of light and $G$ is Newton's gravitational constant).
If two more equations with the same three unknown variables, $M_p$, $M_c$ and $\sin i$, become available, it is possible  to determine the masses of the pulsar and its companion. 
These equations can be found within the so-called Post-Keplerian (PK) formalism where a set of relativistic additions to the classical Keplerian parameters can be parameterized in a theory-independent  
way in various timing models. 

If we assume GR to be the correct  theory of gravity, then these Post-Keplerian (PK) parameters  (such as the orbital period derivative $\dot{P_b}$, the advance of periastron $\dot{\omega}$, 
the gravitational redshift $\gamma$ and the range  \textit{r} and shape \textit{s} of the Shapiro delay) relate the component masses and the Keplerian parameters (see \citealt{TW1982}), 
thus, providing extra equations complementing
Eq.~\ref{eq:mass_function}. If two PK parameters
are available, we can determine the component masses,
if more PK parameters are available we can test the self-consistency
of GR and other theories of gravity.
  
The total mass of the system can be obtained from the
measurement of the periastron advance, $\dot{\omega}$,  according to:
\begin{equation}
M_{\rm tot}\,=\,\frac{1}{T_{\odot}} \left[ \frac{\dot{\omega}}{3} (1- e^2)\right]^{\frac{3}{2}} \left( \frac{P_{\mathrm{b}}}{2\pi}\right)^{\frac{5}{2}}.
\label{eq:M}
\end{equation}

The component masses come from the measurement of the Shapiro delay.
In the DD parameterization \citep{dd85,dd86}, this is described by the
``Range'' $r\,=\,T_{\odot}\, M_c$ and ``Shape" $s = \sin i$ parameters.
In DDH parametrization \citep{fw10}, we have the orthometric ratio and amplitude,
respectively:  
\begin{equation}
\varsigma\,=\, \frac{\sin i}{1+\cos i}, \:   h_3\, = \,r\,\varsigma^{3}.
\label{eq:h3stig}
\end{equation}
The latter parameterization has the advantage of a smaller correlation
between the parameters and also better describes the regions of
the $M_c - \sin i$ plane where the parameters of the system are. For this
reason, when other PK effects are also known, then this parameterization
provides a better test of GR.

\subsubsection{PSR J2045+3633}
\label{sec:analysis_2045}

For PSR J2045+3633 we have used the DDH model to measure the PK parameters {$\dot{\omega}$, $h_3$ and $\varsigma$}.
From the measurement of $\dot{\omega}$ we derived the total mass  $M_{\rm tot}\, = \, 2.28(45)\, M_{\odot}$ using Eq.~\ref{eq:M}. 

We can estimate approximately the masses of the individual components
from the intersection of the $h_3$ and $\dot{\omega}$
curves\footnote{Note that this is not
possible in the $r$-$s$ parameterization, where constraints from
the measurement of $\sin i$ and $M_c$ are too wide to give any useful
mass constraints.} in Fig.~\ref{figure:2045_mass}.
To do this robustly, we performed a Bayesian $\chi^2$
analysis in the $M_c - \cos i$ plane in the fashion described
in \citet{Splaver02}.
For each point in this  plane, we calculated the Shapiro delay parameters
and the rate of advance of periastron
using the specifications of general relativity. Keeping these fixed, we
fitted for the spin, astrometric and Keplerian parameters 
tracking the post-fit  $\chi^2$ (see Fig.~\ref{figure:2045_mass}).  From this $\chi^2$ map we derived a 2-D probability distribution 
function (pdf) that was then translated into a 2-D pdf in the $M_c - M_p$ plane using Eq.~\ref{eq:mass_function}. 
We then marginalize the 2-D pdfs to derive 1-D pdfs for $M_c$, $\cos i$
and $M_p$. 

\begin{figure*}
\includegraphics[width=6in,scale = 0.6]{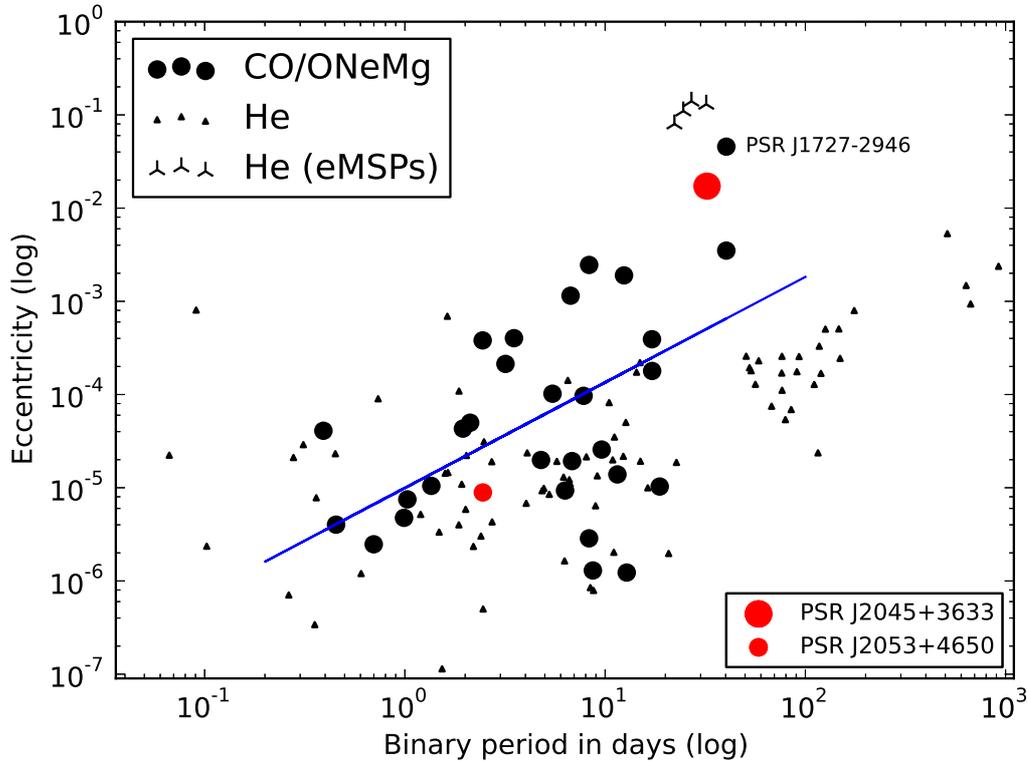}
\caption[P_b-ecc diagram]{Location of the two new MSPs on the  $P_b-ecc$ diagram for the  population of binary pulsars with WD companions and spin periods < 100 ms.
The blue line is a linear regression fit for IMBPs (CO/ONeMg WD companions).}
\label{fig:Ecc_PB}
\end{figure*}

As can be seen from Table~\ref{table:parameters} and Fig.~\ref{figure:2045_mass}, the current precision of the Shapiro delay measurement, even combined with $\dot{\omega}$, does not allow us to constrain the masses precisely. 
This is a consequence of the orbit not being highly inclined, with  $\,i =\, 62\substack{+ 5 \\ -6}^{\circ}$. 
The best-fit mass values within $1\sigma$-band are: $M_p\, = \,1.33\substack{+ 0.30 \\ -0.28}\, M_{\odot}$, $M_c\, = \,0.94\substack{+ 0.14 \\ -0.13}\, M_{\odot}$.
The mass of the pulsar is not yet precise enough for any conclusions,
however, it is clear that the companion is either a heavy CO or ONeMg WD,
as implied by the large mass function of the system.

Continued timing with a special focus on observations near the superior conjunction will greatly improve the precision of the Shapiro delay detection, but even
greater improvements will arise from the fast-improving measurement
of $\dot{\omega}$, for which the uncertainty is proportional to $T^{-3/2}$
(where $T$ is the temporal span of the timing data).

Given the large orbital period of PSR~J2045+3633 ($P_b=32.3\;{\rm days}$), this system probably did not evolve via CE 
(if so, the envelope of the WD progenitor star must have been very weakly bound at the onset of the CE because of the small degree of orbital in-spiral).
This system might instead have formed by stable Case~B RLO  in an IMXB system \citep{tvs00}. It is worth noting that the orbital configuration is also compatible with
 theoretical modelling of IMXBs producing pulsars with massive WDs. However, the large orbital period could possibly
be accounted for by changing the assumptions of the specific orbital angular momentum carried away by the material lost from the system during RLO.

\subsubsection{PSR J2053+4650}
\label{sec:analysis_2053}

For this system, the high orbital inclination and large companion mass
yield a strong signature of the Shapiro delay (Fig.~\ref{fig:resids_2053}) allowing mass measurements from this effect alone.
We fitted the timing data with both DD and DDH models. The masses derived from the range and shape parameters measured within the DD model are consistent with the values
obtained from a Bayesian analysis (Fig.~\ref{figure:2053_mass}) similar to that described in Section~\ref{sec:analysis_2045} performed within the DDH model: $M_p\,= \,1.40\substack{+ 0.21 \\ -0.18}\, M_{\odot}$, 
$M_c\, = \,0.86\substack{+ 0.07 \\ -0.06}\, M_{\odot}$.
The companion is most likely a heavy CO~WD, or possibly an ONeMg~WD.
For this system, although we will not be able to measure any other PK
paramaters in the near future (a couple of years); it is possible that $\dot{\omega}$
might be measurable in the distant future (tens of years).

The combination of a short orbital period ($P_b=2.45\;{\rm days}$) and a massive WD companion suggests that this system
formed via  CE evolution from an IMXB with a donor star on the asymptotic giant branch, e.g.\ following the scenario suggested by \citet{vdh94}.

\subsection{Eccentricity of PSR J2045+3633}

For the majority of known IMBPs observed, eccentricities range from $10^{-6}$ to $10^{-3}$ (see Fig.~\ref{fig:Ecc_PB}). This is certainly the
case for one of our new binary systems, PSR~J2053+4650.
However, some systems have larger eccentricities, one of them being 
PSR J0621+1002 \citep{Splaver02} which has $e\, = \, 0.00245744(5)$.
In 2015, a presumed IMBP, PSR J1727$-$2946, was discovered  with an orbital period
of 40 days and an eccentricity of $e \, = 0.04562943(16)$
 \citep[see][]{lem15}, even larger than that of PSR~J2045+3633 ($e\, = \, 0.01721244(5)$).

Looking at Fig.~\ref{fig:Ecc_PB},  based on the data from the ATNF catalogue\footnote{http://www.atnf.csiro.au/people/pulsar/psrcat/; \cite{mhth05}}, we could in principle say that the relatively high
orbital eccentricities of PSR~J2045+3633 and~PSR J1727$-$2946
are the result of a smooth trend of increasing orbital eccentricity
with orbital period among the systems with massive WD companions.
Indeed, the correlation shown as a blue line in Fig.~\ref{fig:Ecc_PB} was obtained as a result of a linear regression fit to the observed data (in logarithmic scale)
with  $R^2\,=\,0.515$, $p\,=\,0.003$ and $stderr\,=\,0.35$. This low significance for the regression is not surprising if we take into account the large
spread in eccentricities for a given orbital period. 

Given the three different formation channels proposed for IMBPs (see Section~1), the scatter
in Fig.~\ref{fig:Ecc_PB} could simply reflect different origins of these systems.

It is interesting to note that the orbital periods of PSR~J2045+3633 and PSR~J1727$-$2946 are close to those of the 
anomalously eccentric MSP--He~WD systems (eMSPs): PSR J1946+3417 \citep{bck+13, 2017MNRAS.465.1711B}, PSR J2234$+$0611 \citep{Deneva_2013,Antoniadis_2016_ArXiv}, PSR J1950$+$2414 \citep{Knispel_2015} 
and PSR~J0955$-$6150 \citep{Camilo_2015}. Very few (if any) circular systems are observed in this orbital period range
\citep[sometimes dubbed the ``Camilo gap'' first noticed in ][]{cam96}. Although we cannot see any common physical mechanism, it is possible that for 
IMBPs there might also be an eccentricity anomaly at these orbital periods.
This would be very surprising: all of the scenarios put forward to explain the eccentric MSP--He~WD systems
with orbital periods between 22 and 32 days \citep{ft14,antoniadis14,jiang15}
make those predictions for low-mass He~WD companions only.

Further discoveries of IMBPs with these orbital periods and larger will be necessary for determining whether
IMBPs follow a smooth trend in the $P_b-ecc$ diagram, or whether there possibly is a universal eccentricity anomaly for all
binary pulsars with orbital periods between 20 and 40 days.

\section{Summary and conclusions}

We have presented two  binary MSPs discovered in the Northern High Time Resolution Universe pulsar survey. As shown by the timing solutions, 
we have added two new members to the population of intermediate-mass binary pulsars. While PSR J2053+4650 is a standard representative of this population, PSR J2045+3633,
with its relatively large eccentricity $e\, = \, 0.0172$, appears to be atypical (there are only a few other systems with eccentricities of the same order) and especially interesting for studying stellar evolution. 

Both systems are promising for precise mass measurements of their components. Current constraints on the pulsar masses  are  $1.33\substack{+ 0.30 \\ -0.28}\, M_{\odot}$ for PSR J2045+3633 and  $1.40\substack{+ 0.21 \\ -0.18}\, M_{\odot}$
for PSR J2053+4650 where the  median values are in agreement with the assumption that the masses of mildly recycled pulsars should be close to their birth values ${\sim}1.35\, M_{\odot}$ since they accreted little matter 
from their massive companions. 
The precision of measurements will be improved with further timing. Moreover, in the case of PSR J2045+3633, three post-Keplerian parameters (the periastron advance and two Shapiro delay parameters) can be meausured 
with high precision in the near future providing additional constraints.  The low rms of timing residuals and sharp  profiles of both pulsars suggest that they may be  useful for pulsar timing arrays.

\section*{Acknowledgements}

The Arecibo Observatory is operated by SRI International under a cooperative agreement 
with the National Science Foundation (AST-1100968), and in alliance with Ana G. Méndez-Universidad Metropolitana, and the Universities Space Research Association. The part of this work is based on 
observations with the 100-m telescope of the MPIfR (Max-Planck-Institut für Radioastronomie) at Effelsberg. The observations using the Lovell Telescope at Jodrell Bank are supported by a consolidated
grant from the STFC in the UK. The Nançay radio observatory is operated by the Paris Observatory, associated with the French Centre National de la Recherche Scientifique (CNRS). We also acknowledge financial support 
from the `Gravitation, R\'{e}f\'{e}rences, Astronomie, M\'{e}trologie' (GRAM) national programme of CNRS/INSU, France.

\bibliographystyle{mnras}
\bibliography{MSPs_revision}

\bsp	
\label{lastpage}
\end{document}